\documentclass{revtex4}

\begin{document}
\title{Some basics of $su(1,1)$}
\author{Marcel Novaes}

\affiliation{Instituto de F\'{i}sica ``Gleb Wataghin",
Universidade Estadual de Campinas, 13083-970 Campinas-SP, Brazil}

\begin{abstract}
A basic introduction to the $su(1,1)$ algebra is presented, in
which we discuss the relation with canonical transformations, the
realization in terms of quantized radiation field modes and
coherent states. Instead of going into details of these topics, we
rather emphasize the existing connections between them. We discuss
two parametrizations of the coherent states manifold
$SU(1,1)/U(1)$: as the Poincar{\'e} disk in the complex plane and
as the pseudosphere (a sphere in a Minkowskian space), and show
that it is a natural phase space for quantum systems with
$SU(1,1)$ symmetry.

\end{abstract}

\maketitle

\section{Introduction}

The $su(1,1)\sim sp(2,R) \sim so(2,1)$ algebra is defined by the
commutation relations
\begin{equation}\label{comut}
\lbrack K_1,K_2\rbrack=-iK_0, \quad \lbrack K_0,K_1\rbrack=iK_2,
\quad \lbrack K_2,K_0\rbrack=iK_1,
\end{equation}
and it appears naturally in a wide variety of physical problems. A
realization in terms of one-variable differential operators,
\begin{equation}\label{dif}
K_1=\frac{d^2}{dy^2}+\frac{a}{y^2}+\frac{y^2}{16}, \quad
K_2=\frac{-i}{2}\left(y\frac{d}{dy}+\frac{1}{2}\right), \quad
K_0=\frac{d^2}{dy^2}+\frac{a}{y^2}-\frac{y^2}{16},
\end{equation}
for example, allows any ODE of the kind
\begin{equation}
\left(\frac{d^2}{dy^2}+\frac{a}{y^2}+by^2+c\right)f(y)=0
\end{equation}
to be expressed as a $su(1,1)$ element \cite{wybourne1974}. The
radial part of the hydrogen atom and of the $3D$ harmonic
oscillator, and also the Morse potential fall into this category,
and the analytical solution of these systems is actually due to
their high degree of symmetry. In fact, the close relation between
the concepts of symmetry, invariance, degeneracy and integrability
is of great importance to all areas of physics \cite{gilmore1994}.

Just like for $su(2)$, we can choose a different basis
\begin{equation}
K_\pm=(K_1\pm iK_2),
\end{equation}
in which case the commutation relations become
\begin{equation}
\lbrack K_0,K_\pm\rbrack=\pm K_\pm, \quad \lbrack
K_+,K_-\rbrack=-2K_0.
\end{equation}
Note the difference in sign with respect to $su(2)$. The Casimir
operator, the analog of total angular momentum, is given by
\begin{equation}
C=K_0^2-K_1^2-K_2^2=K_0^2-\frac{1}{2}(K_+K_-+K_-K_+).
\end{equation}
This operator commutes with all of the $K$'s.

Since the group $SU(1,1)$ is non-compact, all its unitary
irreducible representations are infinite-dimensional. Basis
vectors $|k,m\rangle$ in the space where the representation acts
are taken as simultaneous eigenvectors of $K_0$ and $C$:
\begin{eqnarray}
C|k,m\rangle=k(k-1)|k,m\rangle, \\
K_0|k,m\rangle=(k+m)|k,m\rangle \label{spec},
\end{eqnarray}
where the real number $k>0$ is called the Bargmann index and $m$
can be any nonnegative integer (we consider only the positive
discrete series). All states can be obtained from the lowest state
$|k,0\rangle$ by the action of the "raising" operator $K_+$
according to
\begin{equation}
|k,m\rangle=\sqrt{\frac{\Gamma (2k)}{m!\Gamma
(2k+m)}}(K_+)^m|k,0\rangle.
\end{equation}

\section{Energy levels of the hydrogen atom}
The hydrogen atom, as well as the Kepler problem, has a high
degree of symmetry, related to the particular form of the
potential. This symmetry is reflected in the conservation of the
Laplace-Runge-Lenz vector, and leads to a large symmetry group,
$SO(4,2)$. Here we restrict ourselves to the radial part of this
problem, as an example of the applicability of group theory to
quantum mechanics and of $su(1,1)$ in particular.  For more
complete treatments see \cite{wybourne1974,gilmore1994}. The
radial part of the Schr\"odinger equation for the hydrogen atom is
\begin{equation}
\left(
\frac{d^2}{dr^2}+\frac{2}{r}\frac{d}{dr}-\frac{2Z}{r}-\frac{l(l+1)}{r^2}+2E\right)R(r)=0.
\end{equation}
If we make $r=y^2$ and $R(r)=y^{-3/2}Y(y)$ we have
\begin{equation}
\left(
\frac{d^2}{dy^2}-\frac{4l(l+1)-3/4}{y^2}+8Ey^2-8Z\right)Y(y)=0,
\end{equation}
and, as already noted in the introduction, this can be written in
terms of the $su(1,1)$ generators (\ref{dif}). A little algebra
gives
\begin{equation}
\left[ (\frac{1}{2}-64E)K_0+(\frac{1}{2}+64E)K_1-8Z\right] Y(y)=0,
\end{equation}
and the Casimir reduces to $C=l(l+1)$, which gives $k=l+1$.

Using the transformation equations
\begin{eqnarray}
e^{-i\theta K_2}K_0e^{i\theta K_2}=K_0 \cosh \theta+K_1 \sinh
\theta \\
e^{-i\theta K_2}K_1e^{i\theta K_2}=K_0 \sinh \theta+K_1 \cosh
\theta
\end{eqnarray}
we can choose
\begin{equation}
\tanh \theta=\frac{64E+1/2}{64E-1/2}
\end{equation}
in order to obtain
\begin{equation}
K_0\tilde{Y}(y)=\frac{Z}{\sqrt{-2E}}\tilde{Y}(y),
\end{equation}
where $\tilde{Y}(y)=e^{-i\theta K_2}Y(y)$. Since we know the
spectrum of $K_0$ from (\ref{spec}) we can conclude that the
energy levels are given by
\begin{equation}
E_n=-\frac{Z}{2n^2},\quad n=m+l+1.
\end{equation}

\section{Relation with $Sp(2,R)$}

A system with $n$ degrees of freedom, be it classical or quantum,
always has $Sp(2n,R)$ as a symmetry group. Classical mechanics
takes place in a real manifold, and the equations of motion are
given by Poisson brackets ($i,j=1..N$)
\begin{equation}\label{PP}
\{q_i,p_j\}=\delta_{ij}.
\end{equation}
Quantum mechanics takes place in a complex Hilbert space, and the
dynamics is determined by the canonical commutation relations
($i,j=1..N$)
\begin{equation}\label{CC}
[\hat{q}_i,\hat{p}_j]=i\hbar\delta_{ij}.
\end{equation}
These relations can also be written in the form (now $i,j=1..2N$)
\begin{eqnarray}\label{Rel}
\{\xi_i,\xi_j\}&=&J_{ij},\\
\lbrack \hat{\xi}_i,\hat{\xi}_j\rbrack&=&i\hbar J_{ij},
\end{eqnarray}
where $\xi=(q_1,...,q_N,p_1,...,p_N)^T$, $\hat{\xi}_i$ is the
hermitian operator corresponding to $\xi_i$ and $J$ is the
$2N\times 2N$ matrix given by
\begin{equation}
J=\left(
\begin{array}{cc}
0  & 1 \\
-1 & 0
\end{array}
\right).
\end{equation}

The symplectic group $Sp(2N,R)$ (in its defining representation)
is composed by all real linear transformations that preserve the
structure of relations (\ref{Rel}). It is easy to see that
therefore
\begin{equation}
Sp(2N,R)=\{S|SJS^T=J\}.
\end{equation}
For a far more extended and detailed discussion, see
\cite{pjp45a1995}

For a classical system with only one degree of freedom, such
canonical transformations are generated by the vector fields
\cite{job2aw2000}
\begin{equation}
\{ -q\frac{\partial}{\partial p}+p\frac{\partial}{\partial
q}=2iK_0, \quad -q\frac{\partial}{\partial
p}-p\frac{\partial}{\partial q}=2iK_1, \quad
-q\frac{\partial}{\partial q}+p\frac{\partial}{\partial
p}=2iK_2,\}.
\end{equation}
It is easy to see that these operators have the same commutation
relations as the $su(1,1)$ algebra (\ref{comut}).

Note that the symplectic groups $Sp(2n,R)$ are non-compact, and
therefore any finite dimensional representation must be
nonunitary. In the quantum case, that means that the matrices $S$
implementing the transformations
\begin{equation}\label{cond}
\hat{\xi}'_j=S_{ij}\hat{\xi}_i,
\end{equation}
such that $\lbrack \hat{\xi}'_i,\hat{\xi}'_j\rbrack=i\hbar
J_{ij}$, are nonunitary (a $2\times 2$ nonunitary representation
of $su(1,1)$ exists for example in terms of Pauli matrices,
$K_1=\frac{i}{2}\sigma_2,K_2=-\frac{i}{2}\sigma_1,K_0=\frac{1}{2}\sigma_3$).
However, since all $\hat{\xi}_i$ and all $\hat{\xi}'_j$ are
hermitian and irreducible, by the Stone-von-Neumann theorem
\cite{pjp45a1995,jordan1974}there exists an operator $U(S)$ that
acts unitarily on the infinite dimensional Hilbert space of pure
quantum states (Fock space). If we now see $\hat{\xi}_i$ and
$\hat{\xi}'_i$ as (infinite dimensional) matrices, then $U(S)$ is
such that $\hat{\xi}'_i=U(S)\hat{\xi}_iU(S)^{-1}$. Finding this
unitary operator in practice is in general a nontrivial task.

\section{optics}
\subsection{one-mode realization}

We know the radiation field can be described by bosonic operators
$a$ and $a^\dag$. If we form the quadratic combinations
\begin{equation}
K_+=\frac{1}{2}(a^\dag)^2, \quad K_-=\frac{1}{2}a^2, \quad
K_0=\frac{1}{4}(aa^\dag+a^\dag a)
\end{equation}
we obtain a realization of the $su(1,1)$ algebra. In this case the
Casimir operator reduces identically to
\begin{equation}
C=k(k-1)=-\frac{3}{16},
\end{equation}
which corresponds to $k=1/4$ or $k=3/4$. It is not difficult to
see that the states
\begin{equation}
|n\rangle=\frac{(a^\dag)^n}{\sqrt{n!}}|0\rangle
\end{equation}
with even $n$ form a basis for the unitary representation with
$k=1/4$, while the states with odd $n$ form a basis for the case
$k=3/4$.

The unitary operator
\begin{equation}\label{1msq}
S(\xi)=\exp\left(\frac{1}{2}\xi^\ast a^2 -\frac{1}{2}\xi
(a^\dag)^2\right)=\exp(\xi^\ast K_--\xi K_+)
\end{equation}
is called the squeeze operator in quantum optics, and is
associated with degenerate parametric amplification
\cite{scully1999}. There is also the displacement operator
\begin{equation}
D(\alpha)=\exp\left(\alpha a^\dag -\alpha^\ast a\right ),
\end{equation}
which acts on the vacuum state $|0\rangle$ to generate the
coherent state
\begin{equation}
|\alpha\rangle=D(\alpha)|0\rangle=e^{-|\alpha|^2/2}\sum_{n=o}^{\infty}
\frac{\alpha^n}{\sqrt{n!}}|n\rangle.
\end{equation}
Action of $S(\xi)$ on a coherent state gives a squeezed coherent
state, $|\alpha,\xi\rangle=S(\xi)|\alpha\rangle$.

\subsection{two-mode realization}

It is also possible to introduce a two-mode realization of the
algebra $su(1,1)$. This is done by defining the generators
\begin{equation}
K_+=a^\dag b^\dag, \quad K_-=ab, \quad K_0=\frac{1}{2}(a^\dag
a+b^\dag b+1).
\end{equation}
In this case the Casimir operator is given by
$C=\frac{1}{4}(a^\dag a-b^\dag b)^2-\frac{1}{4}$. If we introduce
the usual two-mode basis $|n,m\rangle$ then the states
$|n+n_0,n\rangle$ with fixed $n_0$ form a basis for the
representation of $su(1,1)$ in which $k=(|n_0|+1)/2$. A charged
particle in a magnetic field can also be described by this
formalism \cite{jpa36mn2003}.

The unitary operator
\begin{equation}\label{2msq}
S_2(\xi)=\exp\left(\xi^\ast ab -\xi a^\dag
b^\dag\right)=\exp(\xi^\ast K_--\xi K_+)
\end{equation}
is called the two-mode squeeze operator \cite{scully1999}, or
down-converter. When we consider the other quadratic combinations
($\{ a^\dag b, (a^\dag)^2, (b^\dag)^2, a^\dag a-b^\dag b\}$ and
their hermitian adjoint) we have the algebra $sp(4,R)$, of which
$sp(2,R)\sim su(1,1)$ is a subalgebra. More detailed discussions
about group theory and optics can be found for example in
\cite{pjp45a1995,job2aw2000,braunstein1999}.

\section{coherent states}

Normalized coherent states can be defined for a general unitary
irreducible representation of $su(1,1)$ as \cite{perelomov1986}
\begin{equation}
|z,k\rangle=(1-|z|^2)^k\sum_{m=0}^{\infty}\sqrt{\frac{\Gamma
(2k+m)}{m!\Gamma (2k)}}z^m|k,m\rangle,
\end{equation}
where $z$ is a complex number inside the unit disk,
$D=\{z,|z|<1\}$. Similar to the usual coherent states, they can be
obtained from the lowest state by the action of a displacement
operator:
\begin{equation}\label{ztau}
|z,k\rangle=\exp(\zeta K_+-\zeta^\ast K_-)|k,0\rangle, \quad
z=\frac{\zeta}{|\zeta|}\tanh|\zeta|.
\end{equation}
From (\ref{2msq}) we see that $su(1,1)$ coherent states are
actually the result of a two-mode squeezing upon a Fock state of
the kind $|n_0,0\rangle$. On the other hand, from the one-mode
realization (\ref{1msq}) they can be regarded as squeezed vacuum
states.

These states are not orthogonal,
\begin{equation}
\langle
z_1,k|z_2,k\rangle=\frac{(1-|z_1|^2)^k(1-|z_2|^2)^k}{(1-z_1^\ast
z_2)^{2k}}
\end{equation}
and they form an overcomplete set with resolution of unity given
by
\begin{equation}
\int_D \frac{2k-1}{\pi}\frac{dz\wedge dz^\ast
}{(1-|z|^2)^{2}}|z,k\rangle\langle
z,k|=\sum_{m=0}^{\infty}|k,m\rangle\langle k,m|=1 \quad
(k>\frac{1}{2}).
\end{equation}
From the integration measure we see that the coherent states are
parametrized by points in the Poincar{\'e} disk (or
Bolyai-Lobachevsky plane), which we discuss in the next section.
The expectation value for a product of algebra generators like
$K_-^pK_0^qK_+^r$ was presented in \cite{jpa25tl1992} and is given
by
\begin{equation}
\langle z,k|K_-^pK_0^qK_+^r|z,k\rangle=
(1-|z|^2)^{2k}z^{p-r}\sum_{m=0}^{\infty}\frac{\Gamma(m+p+1)\Gamma(m+p+2k)}{m!\Gamma(m+p+1-r)\Gamma(2k)}(m+p+k)^q|z|^{2m}.
\end{equation}
Simple particular cases of this expression are
\begin{equation}
\langle z,k|K_-|z,k\rangle=k\frac{2z}{1-|z|^2}, \quad \langle
z,k|K_0|z,k\rangle=k\frac{1+|z|^2}{1-|z|^2}.
\end{equation}
Moreover, for $k>1/2$ the operator $K_0$ has a diagonal
representation as
\begin{equation}
K_0=\frac{2k-1}{4\pi}\int_D
\frac{d^2z}{(1-|z|^2)^2}(k-1)\left(\frac{1+|z|^2}{1-|z|^2}\right)|z,k\rangle\langle
z,k|.
\end{equation}

Just as usual spin coherent states are parametrized by points on
the space $SU(2)/U(1)\sim S^2$, the two-dimensional spherical
surface, $su(1,1)$ coherent states are parametrized by points on
the space $SU(1,1)/U(1)$, which corresponds to the Poincar{\'e}
disk. This space can also be seen as the two-dimensional upper
sheet of a two-sheet hyperboloid, also known as the pseudosphere.

\section{The Pseudosphere}

The sphere $S^2$ is the set of points equidistant from the origin
in a Euclidian space:
\begin{equation}
S^2=\{(x_1,x_2,x_3)|x_1^2+x_2^2+x_3^2=R^2\}.
\end{equation}
The pseudosphere $H^2$ plays a similar role in a Minkovskian
space, that is, take the space defined by
$\{(y_1,y_2,y_0)|y_1^2+y_2^2-y_0^2=-R^2\}$, which is a two-sheet
hyperboloid that crosses the $y_0$ axis at two points, $\pm R$,
called poles. The pseudosphere, which is a Riemannian space, is
the upper sheet, $y_0>0$. The pseudosphere is related to the
Poincar{\'e} disk by a stereographic projection in the plane
$(y_1,y_2)$, using the point $(0,0,-R)$ as base point. The
relation between the parameters is
\begin{equation}
y_0=R\cosh\tau, \quad y_1=R\sinh\tau\cos\phi, \quad
y_2=R\sinh\tau\sin\phi,
\end{equation}
and
\begin{equation}
z=e^{i\phi}\tanh\frac{\tau}{2}=\frac{y_1+iy_2}{R+y_0}.
\end{equation}

The distance $ds^2=dy_1^2+dy_2^2-dy_0^2$ and the area
$d\mu=\sinh\tau d\tau \wedge d\phi$ become
\begin{eqnarray}
ds^2&=&d\tau^2+\sinh\tau d\phi^2=\frac{dz\cdot
dz^\ast}{(1-|z|^2)^2},\\ d\mu&=&\frac{dz \wedge
dz^\ast}{(1-|z|^2)^2}.
\end{eqnarray}
Note that the metric is conformal, so the actual angles coincide
with Euclidian angles. Geodesics, which are intersections of the
pseudosphere with planes through the origin, become circular arcs
(or diameters) orthogonal to the disk boundary (the non-Euclidian
character of the Poincar{\'e} disk appears in some beautiful
drawings of M.C. Escher, the ``Circle Limit" series
\cite{escher2004}). A very good discussion about the geometry of
the pseudosphere can be found in \cite{pr143nlb1986}, and we
follow this presentation.

In the pseudosphere coordinates the average values of the
$su(1,1)$ generators are very simple:
\begin{equation}
\langle z,k|K_1|z,k\rangle=\frac{k}{R}y_1, \quad \langle
z,k|K_2|z,k\rangle=\frac{k}{R}y_2, \quad  \langle
z,k|K_0|z,k\rangle=\frac{k}{R}y_0.
\end{equation}
From now on we set $R=k=1$.

\subsection{Action of the group}

The symmetry group of the pseudosphere is the group that preserves
the relation $y_1^2+y_2^2-y_0^2=-R^2$, the Lorentz group
$SO(2,1)$. The $so(2,1)$ algebra associated with this group is
isomorphic to the $su(1,1)$ algebra we are studying. All
isometries can be represented by $3\times 3$ matrices $\Lambda$
that are orthogonal with respect to the Minkowski metric $Q={\rm
diag}(1,1,-1)$ (actually we must also impose $\Lambda_{00}>0$ so
that we are restricted to the upper sheet of the hyperboloid), and
they can be generated by $3$ basic types: A) Euclidian rotations,
by an angle $\phi_0$, on the $(y_1,y_2)$ plane; B) Boosts of
rapidity $\tau_0$ along some direction in the $(y_1,y_2)$ plane;
C) Reflections through a plane containing the $y_0$ axis. As
examples, we show a rotation, a boost in the $y_2$ direction and a
reflection through the plane $(y_1,y_0)$:
\begin{equation}
A) \left( \begin{array}{c} \cos\phi_0\\ \sin\phi_0\\0\end{array}
\begin{array}{c}-\sin\phi_0\\ \cos\phi_0\\ 0\end{array}
\begin{array}{c} 0\\0\\1\end{array} \right),
\quad B) \left( \begin{array}{c} 1\\0
\\0\end{array}
\begin{array}{c}0\\ \cosh\tau_0\\ \sinh\tau_0\end{array}
\begin{array}{c} 0\\\sinh\tau_0\\\cosh\tau_0\end{array} \right),
\quad C)\left( \begin{array}{c} 1\\0
\\0\end{array}
\begin{array}{c}0\\ -1\\ 0\end{array}
\begin{array}{c} 0\\0\\1\end{array} \right)
\end{equation}
Incidentally, the geometrical character of the previously used
parameters $(\tau,\phi)$ becomes clear.

Using the complex coordinates of the Poincar{\'e} disk we have
\begin{equation}
R_{\phi_0}(z)=e^{i\phi_0}z
\end{equation}
for rotations,
\begin{equation}
T_{\tau_0,\phi_0}(z)=\frac{(\cosh \tau_0/2)z+e^{i\phi_0}\sinh
\tau_0/2}{(e^{-i\phi_0}\sinh \tau_0/2)z+\cosh \tau_0/2}
\end{equation}
for boosts of rapidity $\tau_0$ in the $\phi_0$ direction and
$S(z)=z^\ast$ for reflections through the $(y_1,y_0)$ plane. We
see that, except for reflections, all isometries can be written as
\begin{equation}
z'=\frac{\alpha z+\beta}{\beta^\ast z+\alpha^\ast}, \quad {\rm
with} \; |\alpha|^2-|\beta|^2=1,
\end{equation}
and if, as usual, we represent these transformations by matrices
$\left( \begin{array}{c} \alpha\\
\beta\end{array}
\begin{array}{c}\beta^\ast\\ \alpha^\ast\end{array}\right)$ there is a realization of the transformation group by
$2\times 2$ matrices, in which
\begin{equation}
R_{\phi_0}=\left( \begin{array}{c} e^{i\phi_0/2}\\
0\end{array}
\begin{array}{c}0\\ e^{-i\phi_0/2}\end{array}
\right), \quad T_{\tau_0,\phi_0}=\left( \begin{array}{c} \cosh \tau_0/2\\
e^{i\phi_0}\sinh \tau_0/2\end{array}
\begin{array}{c}e^{-i\phi_0}\sinh \tau_0/2\\ \cosh \tau_0/2\end{array}
\right).
\end{equation}
This is the basic representation of the group $SU(1,1)$. For other
parametrizations of the pseudosphere, see \cite{pr143nlb1986}.

\subsection{Canonical Coordinates}

We present one last set of coordinates, one that has an important
physical property. Let us first note that if we define
$\mathcal{K}_i=\langle z,k|K_i|z,k\rangle$, then there exists an
operation $\{\cdot,\cdot\}$ such that the commutation relations
\begin{equation}
\lbrack K_1,K_2\rbrack=-iK_0, \quad \lbrack K_0,K_1\rbrack=iK_2,
\quad \lbrack K_2,K_0\rbrack=iK_1
\end{equation}
are exactly mapped to
\begin{equation}
\{ \mathcal{K}_1,\mathcal{K}_2\}=\mathcal{K}_0, \quad \{
\mathcal{K}_0,\mathcal{K}_1\}=-\mathcal{K}_2, \quad
\{\mathcal{K}_2,\mathcal{K}_0\}=-\mathcal{K}_1,
\end{equation}
in agreement with the usual quantization condition
$\{\cdot,\cdot\}\to i[\cdot,\cdot]$. This Poisson Bracket is
written in terms of the Poincar{\'e} disk coordinates as
\begin{equation}
\{f,g\}=\frac{(1-|z|^2)^2}{2ik}\left( \frac{\partial f}{\partial z
}\frac{\partial g}{\partial z^\ast }-\frac{\partial f}{\partial
z^\ast }\frac{\partial g}{\partial z }\right).
\end{equation}

It is possible to define new coordinates $(q,p)$ that are
canonical in the sense that
\begin{equation}
\{f,g\}=\left( \frac{\partial f}{\partial q }\frac{\partial
g}{\partial p}-\frac{\partial f}{\partial p}\frac{\partial
g}{\partial q}\right).
\end{equation}
These coordinates are given by
\begin{equation}
\frac{q+ip}{\sqrt{4k}}=\frac{z}{\sqrt{1-|z|^2}}
\end{equation}
and the classical functions are written in terms of them as
\begin{equation}
\mathcal{K}_1=\frac{q}{2}\sqrt{4k+q^2+p^2}, \quad
\mathcal{K}_2=\frac{p}{2}\sqrt{4k+q^2+p^2}, \quad
\mathcal{K}_0=k+\frac{q^2+p^2}{2}.
\end{equation}
We thus see that there is a natural phase space for quantum
systems that admit $SU(1,1)$ as a symmetry group. Dynamics of
time-dependent systems with this property was examined for example
in \cite{jpa34ab2001}. This phase space can also be used to define
path integrals for $SU(1,1)$ (see \cite{pra39ccg1989,grosche1998}
and references therein), and obtain a semiclassical approximation
to this class of quantum systems.

\section{Summary}
We have presented a very basic introduction to the $su(1,1)$
algebra, discussing the connection with canonical transformations,
the realization in terms of quantized radiation field modes and
coherent states. We have not explored these subjects in their full
detail, but instead we emphasized how they can be related. The
coherent states, for example, can be regarded as one-mode vacuum
squeezed states or as two-mode number squeezed states. The
coherent states manifold $SU(1,1)/U(1)$ was treated as the
Poincar{\'e} disk and as the pseudosphere, and shown to be a
natural phase space for quantum systems with $SU(1,1)$ symmetry.

\end{document}